\documentclass[twocolumn,showpacs,prb,aps]{revtex4}
\usepackage{graphicx}
\usepackage{dcolumn}
\usepackage{amsmath}
\usepackage{color}
\newcommand{\be}{\begin{equation}}
\newcommand{\ee}{\end{equation}}
\newcommand{\bea}{\begin{eqnarray}}
\newcommand{\eea}{\end{eqnarray}}
\newcommand{\bse}{\begin{subequations}}
\newcommand{\ese}{\end{subequations}}

\begin{document}

\title{Antiferromagnetism in EuNiGe$_3$}

\author{R. J. Goetsch}
\author{V. K. Anand}
\author{D. C. Johnston}
\altaffiliation{johnston@ameslab.gov}
\affiliation {Ames Laboratory and Department of Physics and Astronomy, Iowa State University, Ames, Iowa 50011}

%,viewport= 20 20 160 230,clip

\date{\today}

\begin{abstract}

The synthesis and crystallographic and physical properties of polycrystalline EuNiGe$_3$ are reported.  EuNiGe$_3$ crystallizes in the noncentrosymmetric body-centered tetragonal BaNiSn$_3$-type structure (space group $I4mm$), in agreement with previous reports, with the Eu atoms at the corners and body center of the unit cell.  The physical property data consistently demonstrate that this is a metallic system in which Eu spins $S=7/2$ order antiferromagnetically at a temperature $T_{\rm N} = 13.6$~K\@.  Magnetic susceptibility $\chi$ data for $T>T_{\rm N}$ indicate that the Eu atoms have spin~7/2 with $g=2$, that the Ni atoms are nonmagnetic, and that the dominant interactions between the Eu spins are ferromagnetic.  Thus we propose that EuNiGe$_3$ has a collinear A-type antiferromagnetic structure, with the Eu ordered moments in the $ab$-plane aligned ferromagnetically and with the moments in adjacent planes along the $c$-axis aligned antiferromagnetically.  A fit of $\chi(T\leq T_{\rm N})$ by our molecular field theory is consistent with a collinear magnetic structure.  Electrical resistivity $\rho$ data from $T_{\rm N}$ to 350~K are fitted  by the Bloch-Gr\"uneisen model for electron-phonon scattering, yielding a Debye temperature of 265(2)~K\@. A strong decrease in $\rho$ occurs below $T_{\rm N}$ due to loss of spin-disorder scattering.  Heat capacity data at $25~{\rm K} \leq T\leq 300$~K are fitted by the Debye model, yielding the same Debye temperature 268(2)~K as found from $\rho(T)$.  The extracted magnetic heat capacity is consistent with $S=7/2$ and shows that significant short-range dynamical spin correlations occur above $T_{\rm N}$.  The magnetic entropy at $T_{\rm N}=13.6$~K is 83\% of the expected asymptotic high-$T$ value, with the remainder recovered by 30~K\@.

\end{abstract}


\maketitle

\section{INTRODUCTION}

Magnetic susceptibility $\chi$ and magnetization $M$ measurements versus temperature $T$ and applied magnetic field $H$ have been widely used to determine the magnetic properties of materials containing local magnetic moments.\cite{Kittel2005}  Such measurements give valuable information about the magnetic interactions and magnetic structure of a material. In collinear antiferromagnetic (AF) materials, the qualitative $T$ dependence of the anisotropic $\chi$ below the AF ordering (N\'eel) temperature $T_{\rm N}$ is well known. In 1941, Van Vleck calculated using molecular field theory (MFT) the anisotropic $\chi(T\leq T_{\rm N})$ for a collinear AF containing identical crystallographically equivalent spins interacting by Heisenberg exchange, but only for the special case of a two-sublattice bipartite spin lattice, {\it i.e.}, in which the nearest-neighbor spins of a spin on one sublattice (``up spins'') are members of the other sublattice (``down spins'').\cite{VanVleck1941}  He further assumed that spins on one sublattice only interact with the nearest-neighbor spins on the other sublattice and with the same strength.\cite{VanVleck1941}  Due to its limited applicability, few comparisons of experimental $\chi(T\leq T_{\rm N})$ data with these theoretical predictions have been made.

We recently formulated generic predictions using MFT of the anisotropic $\chi(T\leq T_{\rm N})$ of both collinear and planar noncollinear AF structures for Heisenberg spin systems containing identical crystallographically equivalent spins with arbitrary exchange interactions between arbitrary sets of spins.\cite{Johnston2012, Johnston2011} Several comparisons of our theoretical predictions with experimental anisotropic $\chi(T\leq T_{\rm N})$ literature data for single crystals of known collinear and noncollinear AFs were made and reasonable agreement was found.\cite{Johnston2012}  Such comparisons are expected to be most accurate for three-dimensional spin lattices with large spin~$S$, which respectively minimize quantum fluctuations associated with low spin lattice dimensionality and/or low spin that are not taken into account by MFT\@.  The MFT that we formulated is also expected to be most accurate for spin-only ions with angular momentum $L=0$, which minimizes crystalline electric field effects arising from the spin-orbit interaction such as single-ion anisotropy effects.  As discussed in Ref.~\onlinecite{Johnston2012}, the deviation of the MFT prediction from the anisotropic $\chi(T\leq T_{\rm N})$ data for Heisenberg spin systems can be used as a quantitative diagnostic for dynamical spin fluctuations and correlations beyond MFT\@.  Comparisons of $\chi(T)$ of an AF compound with MFT predictions have been used in the past to test for the occurrence of such dynamical short-range spin correlations, but usually only at temperatures above $T_{\rm N}$.

In Ref.~\onlinecite{Johnston2012} we pointed out that the same MFT predictions used to extract information about the magnetic interactions and magnetic structures of AFs from analyses of anisotropic $\chi(T\leq T_{\rm N})$ data for single crystal AFs should also be useful for analyzing the necessarily isotropic $\chi(T\leq T_{\rm N})$ data for polycrystalline AFs.  For example, such measurements can distinguish between collinear and planar noncollinear AF structures, even when multiple collinear AF domains occur.  They can also be used to estimate the wave vector and turn angle between adjacent planes of spins along the helix or cycloid axis of planar helical or cycloidal AF structures.\cite{Yoshimori1959}

We report in this paper $M(H,T)$ and $\chi(T)$ measurements of polycrystalline EuNiGe$_3$ and demonstrate that this compound exhibits long-range AF order below $T_{\rm N}=13.6$~K\@.  We analyze the $\chi(T)$ data at $T\leq T_{\rm N}$ using our new MFT as well as at $T\geq T_{\rm N}$ using the conventional Curie-Weiss law, which is also a MFT prediction. We also report x-ray diffraction measurements of the crystal structure of this material, electrical resistivity $\rho(T)$ measurements fitted by the Bloch-Gr\"uneisen model and heat capacity $C_{\rm p}(T)$ measurements analyzed using the Debye model and correlate the results with the magnetic measurements.

The compound EuNiGe$_3$ crystallizes in the body-centered tetragonal ${\rm BaNiSn_3}$-type structure (space group $I4mm$)  with the Eu atoms at the corner and body-center positions forming a square lattice in the $ab$-plane stacked in a zigzag ABA fashion along the $c$-axis as shown below in Fig.~\ref{fig:structure}(a).\cite{Oniskovets1987, Salamakha1996} No information is available about its physical properties.  Recent measurements of the physical properties of the related compounds $R$NiGe$_3$ ($R=$ Y, Ce--Nd, Sm, Gd--Lu) with the same stoichiometry but with a different base-centered orthorhombic SmNiGe$_3$-type crystal structure (space group $Cmmm$) have been reported, and most of these are found to order antiferromagnetically.\cite{Mun2010}  These compounds also contain a square $R$ sublattice with a different ABBA zigzag stacking sequence, but which is qualitatively similar to the stacked Eu square lattice in EuNiGe$_3$.  Therefore EuNiGe$_3$ also appeared to us to be a candidate for AF ordering as we subsequently confirmed. Also, spin-only Eu$^{+2}$ ions with $S=7/2$ have orbital angular momentum $L=0$, an advantageous property resulting in negligible crystalline electric field effects and a spectroscopic splitting factor $g\approx 2$. 

From analysis of our $\chi(T)$ data for EuNiGe$_3$ at $T\geq T_{\rm N}$ using the Curie-Weiss law, the dominant interactions between the Eu$^{+2}$ spins $S = 7/2$ are ferromagnetic (FM), in spite of the collinear long-range AF order at $T\leq T_{\rm N}$ suggested by our $\chi$ data at $T\leq T_{\rm N}$.  Taking into account the symmetry of the unit cell, we propose that the Eu spins within a tetragonal $ab$-plane interact ferromagnetically but spins in adjacent layers along the $c$-axis interact antiferromagnetically.  We further propose that this set of exchange interactions gives rise to a collinear A-type AF structure in which the Eu ordered moments within a layer are aligned ferromagnetically with respect to each other, but are aligned antiferromagnetically with respect to the moments in the two adjacent planes along the $c$-axis.

For the scheme of magnetic interactions in EuNiGe$_3$ that we propose, the FM interactions between spins within an $ab$-plane act within the same sublattice, and hence the interactions are not consistent with Van Vleck's MFT for $\chi(T<T_{\rm N})$ discussed above, even though both AF structures are collinear.  This means that Van Vleck's prediction for $\chi(T<T_{\rm N})$ is not appropriate for analyzing such data.  Our generic MFT must be used instead.  An analysis of $\chi(T<T_{\rm N})$ for polycrystalline EuNiGe$_3$ cannot determine the orientation of the easy axis of the A-type AF structure with respect to the crystal axes, {\it e.g.}, along the $c$-axis or within the $ab$-plane, although such a determination is possible using $\chi(T)$ data for single-crystal samples.  Future magnetic neutron and/or magnetic x-ray scattering measurements can test our model for the A-type AF structure and also determine the direction of the ordered moments.  On the basis of the analysis of $\chi(T\geq T_{\rm N})$ in terms of the Curie-Weiss law, we obtain estimates of the nearest-neighbor in-plane and out-of-plane Eu-Eu exchange interactions.

The remainder of the paper is organized as follows. In Sec.~\ref{ExpDetails}, we discuss the synthesis of the  polycrystalline EuNiGe$_3$ sample along with details of the measurements. The experimental results, analyses and discussion are presented in Sec.~\ref{Results}.  A summary and our conclusions are given in Sec.~\ref{Conclusion}.

\section{\label{ExpDetails} Experimental Details}

Polycrystalline EuNiGe$_3$ was prepared from the high purity elements.  Eu was obtained from Ames Laboratory and Ni (99.996\%) and Ge (99.9999+\%) from Alfa Aesar. Eu pieces were surrounded by Ni and Ge powders in a pressed pellet. The pellet was placed in a 2~mL alumina crucible and sealed in an evacuated fused silica tube. The sample was heated at 850~$^\circ$C for 30~h followed by a thorough grinding to ensure homogeneity, and then after pelletizing heated in an evacuated fused silica tube at 900~$^\circ$C for $7$~d. All sample handling, except for brief exposures to air to press the pellets and load the silica tubes, was done in a glove box containing high-purity He gas.  A single-phase sample (apart from a trace of elemental Ge) was obtained as established from powder x-ray diffraction measurements described in Section~\ref{sec:structure} below. Growths of single crystals using Sn and NiGe$_3$ fluxes were attempted but were not successful. 

Powder x-ray diffraction (XRD) data were collected using a Rigaku Geigerflex diffractometer with Cu~K$\alpha$ radiation.  Rietveld refinement of the XRD data was accomplished using the FullProf package.\cite{Rodriguez1993}

$M$ measurements versus $H$ and $T$ were carried out using a superconducting quantum interference device (SQUID) magnetometer (Quantum Design, Inc.). A gel cap was used as sample holder and its small diamagnetic magnetization was measured separately and corrected for in the magnetization data for EuNiGe$_3$ that are presented.  We use Gaussian cgs units for the magnetization, magnetic susceptibility and magnetic field throughout, where the magnetic field unit of Tesla, when it appears,  is a unit of convenience (${\rm 1~T \equiv 10\,000~Oe}$).

Heat capacity $C_{\rm p}$ and electrical resistivity $\rho$ measurements were carried out using a Quantum Design Physical Property Measurement System (PPMS). The sample for $C_{\rm p}$ measurements was attached and thermally coupled to the addenda with Apiezon~N grease.  The $\rho$ measurements utilized a four-probe ac technique with the ac-transport option of the PPMS, where a rectangular parallelopiped-shaped sample was cut from the sintered pellet for the measurements using a jeweler's saw.  Platinum electrical leads were attached to the sample using EPO-TEK P1011 silver epoxy and the sample was attached to the resistivity puck with GE~7031 varnish. The $\rho(T)$ measurements were carried out on both cooling and heating to check for hysteresis.

\section{\label{Results} Experimental Results, Analyses and Discussion}

\subsection{\label{sec:structure} Crystal Structure Determination}

\begin{figure}
	\includegraphics[width=3.3in]{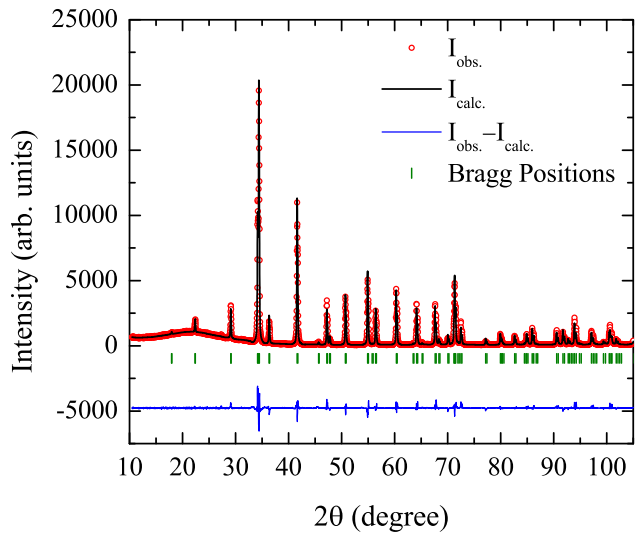}
	\caption{(Color online) Room temperature powder XRD pattern (red circles) of EuNiGe$_3$, Rietveld refinement fit (solid black line), difference profile (lower solid blue line), and positions of Bragg peaks (vertical green bars).}
	\label{fig:XRD}
\end{figure}

\begin{figure}
	\includegraphics[width=3in]{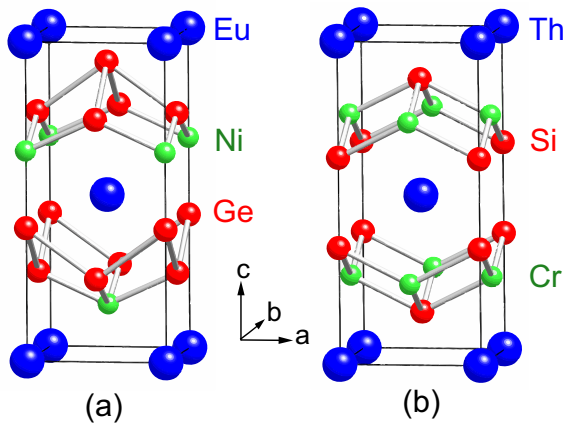}
	\caption{(Color online) Comparison of the structure of (a) EuNiGe$_3$ [BaNiSn$_3$-type (space group {\it I}4{\it mm})] with (b) the ThCr$_2$Si$_2$-type structure (space group {\it I}4/{\it mmm}).}
	\label{fig:structure}
\end{figure}

 \begin{table}
 \caption{\label{table:XRD} Crystal data for EuNiGe$_3$ at room temperature (BaNiSn$_3$-type structure: space group $I4mm$). Refined lattice parameters and unit cell volume are $a=b=4.3366(1)$~\AA, $c=9.8802(2)$~\AA\ and $V_{\rm cell}=185.81(1)$~\AA$^3$. Listed are the Wyckoff atomic position and the atomic coordinates $x$, $y$, and $z$ of each atom type. The goodness-of-fit parameters obtained are $R_{\rm p}=7.41$\%, $R_{\rm wp}=10.4$\%, and $\chi^2=5.27$.}
 \begin{ruledtabular}		
		\begin{tabular}{l c c c c}
		Atom & Wyckoff position & $x$ & $y$ & $z$  	\\ \hline
		Eu 	& 2{\it a}	& 0 & 0 	& 0.0028(3)		\\
		Ni 	& 2{\it a} 	& 0 & 0 	& 0.6581(4) 	\\
		Ge1	& 4{\it b} 	& 0 & 1/2 & 0.2582(4) 	\\
		Ge2	& 2{\it a} 	& 0 & 0 	& 0.4163(4)  	\\
		\end{tabular}
 \end{ruledtabular}
\end{table}

The crystal structure of EuNiGe$_3$ reported in Ref.~\onlinecite{Oniskovets1987} was used as the starting point for the Rietveld refinement of our powder XRD data. We also observed very weak peaks from an impurity phase with the strongest peaks at diffraction angles $2\theta=27.3^\circ$ and $45.3^\circ$. These correspond to the strongest peaks of pure Ge, indicating the presence of a trace amount of elemental Ge in our sample. During refinement of the XRD data for the EuNiGe$_3$ phase, the thermal parameters of the atoms were fixed at zero. In the final refinement the occupancies of the atoms were fixed at the stoichiometric values because no significant difference in the goodness of fit was obtained when the occupancies were allowed to vary.

A satisfactory Rietveld refinement of the powder XRD data for EuNiGe$_3$ was obtained assuming the body-centered-tetragonal BaNiSn$_3$-type structure (space group $I4mm$) as shown in Fig.~\ref{fig:XRD}, with parameters listed in Table~\ref{table:XRD}\@.  This structure and the common ThCr$_2$Si$_2$-type structure are both derivatives of the BaAl$_4$-type structure.\cite{Parthe1983} A comparison between the EuNiGe$_3$ and ThCr$_2$Si$_2$ structures is shown in Fig.~\ref{fig:structure}. The Eu and Th positions are identical in the two structures.  The ThCr$_2$Si$_2$-type structure is centrosymmetric whereas the BaNiSn$_3$-type structure is not. In the BaNiSn$_3$-type structure, the transition metal square lattice in the $ab$-plane is rotated by $45^\circ$  with respect to the $ab$-plane Cr square-lattice in ThCr$_2$Si$_2$, and the ordering of the Si or Ge layers and the transition metal layers along the $c$-axis is different. 

The refined lattice parameters for EuNiGe$_3$ in the caption of Table~\ref{table:XRD} can be compared with the reported values $a=4.737(2)$~\AA\ and $c=9.891(3)$~\AA.\cite{Oniskovets1987}  Our $a$-axis parameter is much smaller by 0.400~\AA\ than the reported value, which we therefore assume is due to a typographical error in Ref.~\onlinecite{Oniskovets1987}, but the $c$-axis parameters are nearly the same.   

\subsection{\label{Transport} Electrical Resistivity Measurements}

\begin{figure}
	\includegraphics[width=3.3in]{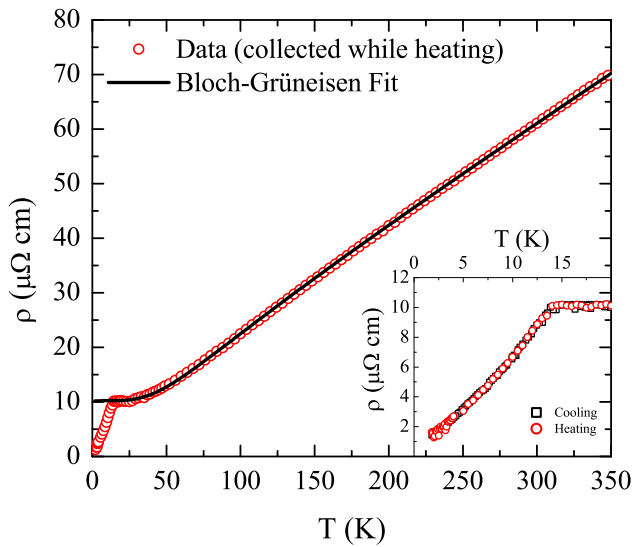}
	\caption{(Color online) Electrical resistivity $\rho$ versus temperature $T$ for EuNiGe$_3$ taken on heating (open red circles). For clarity, only every other data point is plotted.  The black curve is a fit of the data at $T>14$~K by Eq.~(\ref{eq:Gruneisen}).  An extrapolation of the fit to $T=0$ is also shown. Inset: Expanded plot of the $\rho(T)$ data at low~$T$ that were taken on cooling (black squares) and heating (open red circles).}
	\label{fig:Rho}
\end{figure}

The $\rho$ of EuNiGe$_3$ was measured from $T=1.8$ to $350$~K and the data are presented in Fig.~\ref{fig:Rho}. The sample shows a very large residual resistivity ratio ${\rm RRR}=\rho(350\ \text{K})/\rho(1.8\ \text{K})=54.8$ indicating that our polycrystalline sample is of good quality for electronic transport measurements. A strong decrease in $\rho$ at $T<T_{\rm N}$ occurs due to the loss of spin disorder scattering below $T_{\rm N}\approx 13.6$~K (see also below), as shown on expanded scales in the inset of Fig.~\ref{fig:Rho}.  The data in the inset also show no hysteresis between heating and cooling runs, indicating that the AF transition is thermodynamically of second order. Above $T_{\rm N}$ the resistivity due to spin disorder scattering is expected to be constant.\cite{Coles1958} Therefore, the $T$ dependence above $T_{\rm N}$ is due to other electron scattering mechanisms.  Typical mechanisms are electron-electron scattering which leads to a $T^2$ dependence and electron-phonon scattering with or without simultaneous \mbox{Umklapp} scattering.

The Bloch-Gr\"uneisen model predicts the contribution to $\rho(T)$ due to scattering of electrons by longitudinal lattice vibrations in the absence of Umklapp scattering.\cite{Ziman1960,Blatt1968,Goetsch2012}  When additional constant terms are added to account for the residual resistivity ($\rho_0$) and the spin disorder resistivity at $T > T_{\rm N}$ ($\rho_{\rm sd}$), the sum is
\bea
 \rho (T)&=& \rho_0+\rho_{\rm sd}  \nonumber \\
&&+\ 4 \mathcal{R} \left( \frac{T}{\Theta _{\rm{R}}}\right) ^5 \int_0^{\Theta_{\rm R}/T} \hspace{-.2in} \frac{x^5}{(e^x-1)(1-e^{-x})}dx, 
 \label{eq:Gruneisen}
\eea
where $\mathcal{R}$ is a material-dependent prefactor that is independent of $T$ and $\Theta _{\rm{R}}$ is the Debye temperature determined from resistivity measurements. 

To fit our $\rho(T>T_{\rm N})$ data by Eq.~(\ref{eq:Gruneisen}), we utilized a high-accuracy analytic Pad\'e approximant for the Bloch-Gr\"uneisen function in Eq.~(\ref{eq:Gruneisen}) that we formulated recently.\cite{Goetsch2012}  As seen in Fig.~\ref{fig:Rho}, an excellent fit by Eq.~(\ref{eq:Gruneisen}) was obtained to the $\rho(T)$ data for $T\geq T_{\rm N}$. The parameters obtained from the fit are $\rho_0+\rho_{\rm sd} = 10.21(3)\ \mu\Omega$\,cm, $\Theta_{\rm R}=265(2)$~K and $\rho(T=\Theta_{\rm R})= 44.3(2)\ \mu\Omega$\,cm. The quoted statistical errors on the resistivity contributions do not take into account an estimated systematic error of order 10\% arising from uncertainty in the geometric factor and from the porosity and grain boundary scattering of the sintered sample.  The excellent agreement of the temperature dependence of the data with the fit indicates that electron-phonon scattering is the primary scattering mechanism giving rise to the $T$ dependence of $\rho$ for $T>T_{\rm N}$.  This conclusion is supported by the agreement of $\Theta_{\rm R}=265(2)$~K with the Debye temperature $\Theta_{\rm D}=268(2)$~K obtained below by fitting the lattice heat capacity by the Debye model over approximately the same $T$ range.

\subsection{\label{Magnetic} Magnetization and Magnetic Susceptibility Measurements}

\begin{figure}
	\includegraphics[width=3.1in]{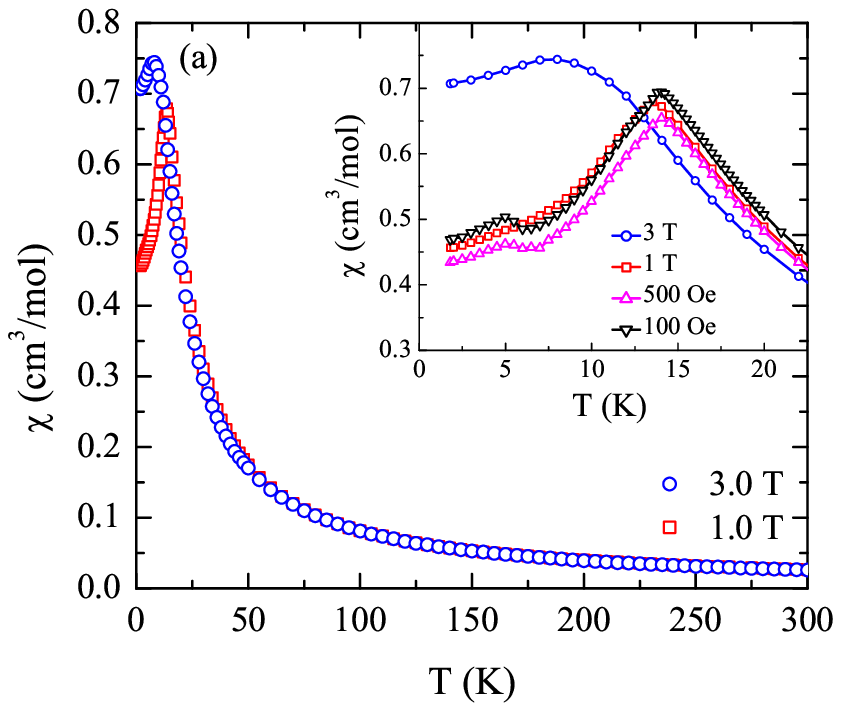}
	\includegraphics[width=3.1in]{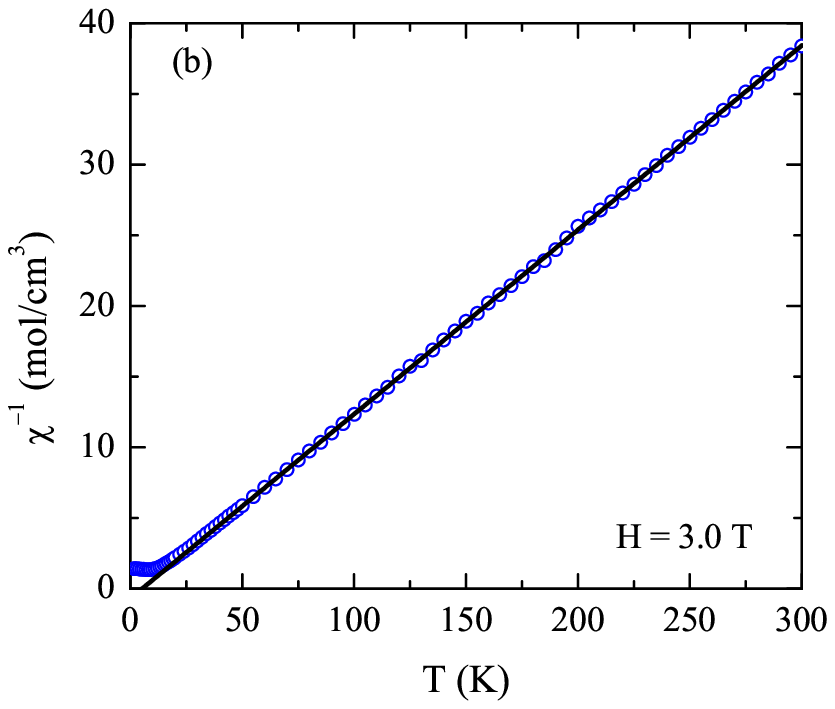}
	\includegraphics[width=3.1in]{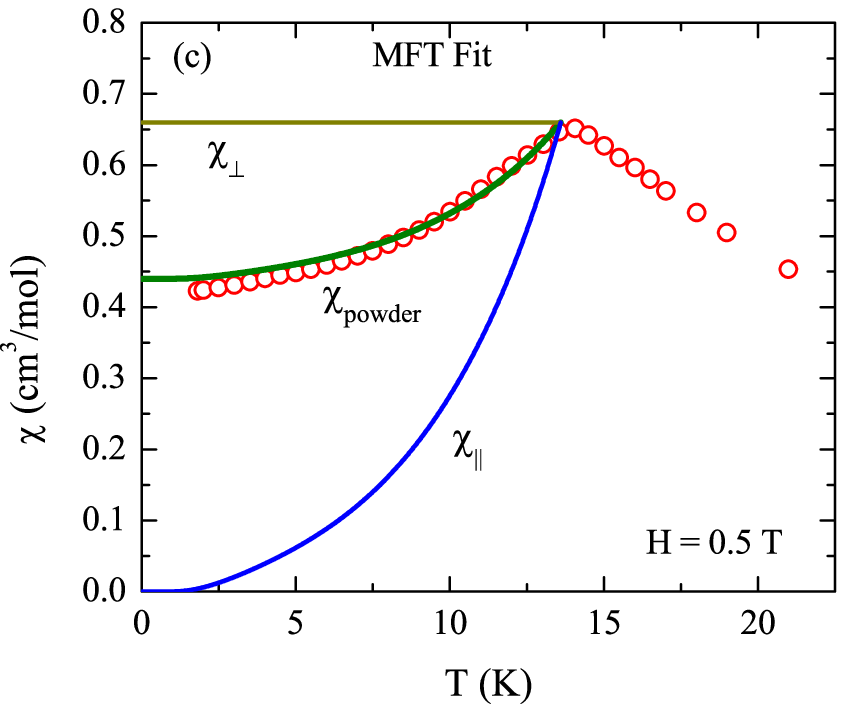}
	\caption{(Color online) Magnetic suceptibility $\chi$ (a) and $\chi^{-1}$ (b) of polycrystalline EuNiGe$_3$ versus temperature $T$.  The inset in~(a) shows expanded plots below 25~K of $\chi(T)$ with $H=100$~Oe to 3~T, and  the straight black line in~(b) is a fit of the data at $T>50$~K by the Curie-Weiss law in Eq.~(\ref{eq:Curie-Weiss}). An extrapolation of the fit to the horizontal axis is also shown.  (c) Expanded plot of $\chi(T)$ at low $T$ taken with $H=0.5$~T\@. The curves are the MFT predictions for $\chi_\perp$, $\chi_\parallel$, and $\chi_{\rm powder}$ versus $T$ from Eqs.~(\ref{Eqs:ChiAnis}) and~(\ref{eq:chi_powder}) compared with the experimental $\chi_{\rm powder}$ data (open red circles).}
	\label{fig:Chi}
\end{figure}

The magnetic susceptibility $\chi \equiv M/H$ versus $T$ of EuNiGe$_3$ was measured at $H = 1$~T and 3~T in the $T$ range 1.8--300~K as shown in Fig.~\ref{fig:Chi}(a).  The inverse susceptibility $\chi^{-1}(T)$ for $H=3$~T is plotted in Fig.~\ref{fig:Chi}(b), and $\chi(T)$ data for $H = 0.5$~T from $T=1.8$ to 25~K are shown in Fig.~\ref{fig:Chi}(c). The sharp peak at $T_{\rm N}\approx 14$~K in the $\chi(T)$ data for $H\leq 1$~T in Figs.~\ref{fig:Chi}(a) and~\ref{fig:Chi}(c) is the signature of AF ordering occurring at this $T$\@.  Another transition at $\approx 5$~K is observed as a sharp cusp for fields $H\lesssim 500$~Oe as shown in the inset of Fig.~\ref{fig:Chi}(a). The nature of this transition is unclear.  There is no evidence of a transition at this $T$ in either the $\rho(T)$ data presented above or the $C_{\rm p}(T)$ data presented below. A similar unexplained feature at about the same temperature was observed in EuPdGe$_3$.\cite{Kaczorowski2012}  The feature cannot be due to a ferromagnetic EuO impurity phase because the Curie temperature of EuO is 69~K.\cite{Liu2012}

At $T> T_{\rm N}$, the molar $\chi$ follows the Curie-Weiss law
\be
\chi(T)=\frac{C}{T-\theta_{\rm p}}
\label{eq:Curie-Weiss}
\ee
as shown by the plot of $\chi^{-1}$ versus~$T$ in Fig.~\ref{fig:Chi}(b) for $H=3$~T, where $C$ is the molar Curie constant and $\theta_{\rm p}$ is the Weiss temperature. These $\chi^{-1}(T)$ data decrease linearly with decreasing $T$ down to $\approx 50$~K, below which short-range AF correlations begin to cause a deviation from the Curie-Weiss law. Thefore we fitted the data in Fig.~\ref{fig:Chi}(b) by Eq.~(\ref{eq:Curie-Weiss}) only from 50 to 300~K\@.  The resulting fit gave $C=7.67(2)$~cm$^3$\,K/mol and $\theta_{\rm p}=5.3(3)$~K\@.  The $C$ value is close to the value $C=7.88$~cm$^3$\,K/mol expected for Eu$^{+2}$ with $S=7/2$ and $g=2$. The agreement of the Curie constant with that expected for Eu$^{+2}$ with $S=7/2$ indicates that the Ni atoms are nonmagnetic.  The conclusion that the Eu atoms have $S=7/2$ agrees with the analysis of the magnetic entropy described in Section~\ref{HC} below.  The ratio $\theta_{\rm p}/T_{\rm N}$ is 
\be
f \equiv \frac{\theta_{\rm p}}{T_{\rm N}} = 0.39(2),
\label{Eq:fDef}
\ee
where we used the precise value $T_{\rm N} = 13.6$~K determined from our heat capacity measurements below.

Using the Heisenberg Hamiltonian ${\cal H} = \sum_{\langle ij\rangle}J_{ij}{\bf S}_i\cdot{\bf S}_j$ for a system comprised of identical crystallographically equivalent spins, where the sum is over distinct pairs of spins, from MFT one can write $\theta_{\rm p}$ and $T_{\rm N}$ in terms of the exchange interactions $J_{ij}$ between spin $i$ and its neighbors~$j$ as\cite{Johnston2012} 
\bse
\label{Eqs:thetaTNJijs}
\bea
\theta_{\rm p} &=& -\frac{S(S+1)}{3k_{\rm B}}\sum_j J_{ij},\label{Eq:thetapSum}\\*
T_{\rm N} &=& -\frac{S(S+1)}{3k_{\rm B}}\sum_jJ_{ij}\cos\phi_{ji},
\eea
\ese
where $k_{\rm B}$ is Boltzmann's constant and $\phi_{ji}$ is the angle between ordered moments~$j$ and~$i$ in the magnetically-ordered state.  There is no restriction on the range of the exchange interactions $J_{ij}$ in Eqs.~(\ref{Eqs:thetaTNJijs}), and these can therefore be nearest-neighbor, next-nearest-neighbor, {\it etc.}, interactions.

From Eq.~(\ref{Eq:thetapSum}), the positive value of $\theta_{\rm p}$ observed for EuNiGe$_3$ indicates that the dominant interactions between the Eu spins are FM (negative), in spite of the long-range AF ordering. In order to simultaneously satisfy these two conditions within the symmetry constraints of the crystal structure, we propose that the dominant FM interactions $J_1$ are between nearest-neighbor Eu spins in the $ab$-plane, with  subdominant AF interactions $J_c$ between nearest-neighbor Eu spins in adjacent layers.  These interactions give rise to an often-observed A-type AF structure in which FM aligned layers of ordered Eu moments in the $ab$-plane are AF aligned with the Eu moments in adjacent Eu planes along the $c$-axis.  Our magnetization data for polycrystalline EuNiGe$_3$ cannot determine the axis along which the ordered moments are aligned, but single-crystal measurements could determine that.  A-type AF structures were reported for the magnetic Co atoms in single-crystal ${\rm CaCo_2As_2}$ with the Co ordered moments aligned along the $c$-axis,\cite{Cheng2012,Ying2012,Lamsal2012} and for the magnetic Eu atoms in single-crystal ${\rm EuFe_2As_2}$ with the Eu ordered moments aligned in the $ab$-plane.\cite{Xiao2009}  Both compounds have the tetragonal ${\rm ThCr_2Si_2}$-type crystal structure shown in Fig.~\ref{fig:structure}(b).

The Weiss and N\'eel temperatures for an A-type AF of stacked square lattices as in Fig.~\ref{fig:structure}(a)  with only nearest-neighbor interactions are given by Eqs.~(\ref{Eqs:thetaTNJijs}) as
\bse
\label{Eqs:JsFromThetapTN}
\bea
\theta_{\rm p} &=& -\frac{S(S+1)}{3k_{\rm B}}(4J_1+8J_c),\quad({\rm A\ type\ AF})\label{Eq:thetapSum2}\\*
T_{\rm N} &=& -\frac{S(S+1)}{3k_{\rm B}}(4J_1-8J_c),
\label{Eq:TNSum2}
\eea
\ese
where for the A-type stacked square lattice AF, by definition one has $\phi_{ji}=0$ for the four nearest-neighbor spin pairs within an $ab$ plane and $\phi_{ji}=180^\circ$ for the eight nearest-neighbor spin pairs between adjacent layers along the $c$-axis.  In the latter case, due to the lack of a horizontal mirror plane through the body-centered Eu site in the noncentrosymmetric crystal structure in Fig.~\ref{fig:structure}(a), the two $J_c$ values from an Eu spin to the four nearest-neighbor Eu spins in each of the two adjacent $ab$-plane layers, respectively, are different. Therefore the derived $J_c$ is an average of the two interplanar interactions.  For a spin lattice consisting of square lattices stacked directly above and below each other, the coefficient of $J_c$ in Eqs.~(\ref{Eqs:JsFromThetapTN}) would have been 2 instead of 8.  From Eqs.~(\ref{Eqs:JsFromThetapTN}) one can solve for the two exchange interactions $J_1$ and $J_c$ in terms of the measured values of $T_{\rm N}$ and $\theta_{\rm p}$, yielding
\bse
\label{Eqs:Jcalcs}
\bea
\frac{J_1}{k_{\rm B}} &=& -\frac{3(T_{\rm N} + \theta_{\rm p})}{8S(S+1)},\quad({\rm A\ type\ AF})\\*
\frac{J_c}{k_{\rm B}} &=& \frac{3(T_{\rm N} - \theta_{\rm p})}{16S(S+1)}.
\eea
\ese
Using $T_{\rm N}=13.6$~K, $\theta_{\rm p}=5.3(3)$~K and $S=7/2$, Eqs.~(\ref{Eqs:Jcalcs}) yield 
\be
\frac{J_1}{k_{\rm B}} = -0.45(1)~{\rm K},\qquad \frac{J_c}{k_{\rm B}} = 0.099(4)~{\rm K}.  
\ee
These results quantitatively confirm the above qualitative deduction based on the positive Weiss temperature that the dominant Eu-Eu exchange interactions in the system are ferromagnetic (negative).  These we deduce to be the $J_1$ interactions between Eu spins within an $ab$-plane layer, whereas the interlayer interactions $J_c$ are antiferromagnetic (positive).  In particular, the dominant $J_i$ ($i=1$ or $c$) value as well as the dominant $z_iJ_i$ value are both negative, where $z_i=4$ and~8 are the   coordination numbers of Eu by Eu for in-plane $J_1$ and out-of-plane $J_c$, respectively.

For a Heisenberg system of identical crystallographically equivalent spins in the absence of magnetocrystalline anisotropy, our MFT predicts the anisotropic temperature dependence of the susceptibility at $T\leq T_{\rm N}$ for a collinear AF such as the A-type AF to be\cite{Johnston2012}
\bse
\label{Eqs:ChiAnis}
\bea
\chi_\perp(T\leq T_{\rm N})&=&\chi(T_{\rm N})\\*
\chi_{\parallel}(t)&=&\left[\frac{1-f}{\tau^*(t)-f}\right] \chi(T_{\rm N})\\*
\tau^*=\frac{(S+1) t}{3 B_{\rm S}^\prime(y_0)},&&\quad y_0= \frac{3 \bar{\mu}_{0}}{(S+1)t},
\eea
with the Brillouin function $B_{S}(y)$ and its derivative $B_S^\prime(y)$ respectively given by
\bea
 B_{S}(y)&=&\frac{1}{2S}\left\{ (2S+1)\coth\left[(2S+1)\frac{y}{2}\right] -\coth\left( \frac{y}{2} \right) \right \},\nonumber \\*
 B_S^\prime(y)&\equiv&\frac{dB_{S}(y)}{dy}\label{eq:Brillouin} \\*
 &=&\frac{{\rm csch}^2(y/2)-(2S+1)^2 {\rm csch}^2[(2S+1)y/2]}{4S},\nonumber
\eea
\ese
where $\parallel$ and $\perp$ refer to the magnetic field applied parallel and perpendicular to the easy axis, respectively, the reduced temperature is $t\equiv T/T_{\rm N}$, and we use the unconventional definition of $B_S(y)$ in Refs.~\onlinecite{Johnston2011} and~\onlinecite{Reif1965}.  The reduced $T$-dependent ordered moment is $\bar{\mu}_0 \equiv \mu_0/\mu_{\rm sat}$ where $\mu_0$ is the magnitude of the ordered moment at $H=0$ and $\mu_{\rm sat} = gS\mu_{\rm B}$ is the saturation moment.  The $\bar{\mu}_0(t)$ is determined by numerically solving the expression $\bar{\mu}_0 = B_S(y_0)$.\cite{Johnston2012,Johnston2011}  Equations~(\ref{Eqs:ChiAnis}) predict that $\chi_\parallel(T=0)=0$ and that the susceptibility is isotropic at $T =  T_{\rm N}$, {\it i.e.}, $\chi_\parallel(T_{\rm N})=\chi_\perp(T_{\rm N}) = \chi(T_{\rm N})$.\cite{Johnston2012}  The $\chi$ follows the Curie-Weiss law in Eq.~(\ref{eq:Curie-Weiss}) and is isotropic for $T>T_{\rm N}$.\cite{Johnston2012,Johnston2011}

In a polycrystalline sample such as ours, it is assumed that the many small crystallites are randomly oriented. Therefore, the $\chi(T)$ can be obtained as the spherical (``powder'') average of the $\chi_\perp$ and $\chi_\parallel$ components to be
\be
\chi_{\rm powder}(T)=\frac{1}{3} [2\chi_\perp(T)+\chi_\parallel(T)].
\label{eq:chi_powder}
\ee
This powder average is the same if multiple A-type AF domains occur.  Using the values $S = 7/2$, $f=0.39$ from Eq.~(\ref{Eq:fDef}) and the observed $\chi(T_{\rm N}) = 0.66(1)\,{\rm cm^3/mol}$ from Fig.~\ref{fig:Chi}(c), the $T$~dependence of the powder average susceptibility for $T\leq T_{\rm N}$ obtained from Eqs.~(\ref{Eqs:ChiAnis}) and~(\ref{eq:chi_powder}) is the green line in Fig.~\ref{fig:Chi}(c) with no adjustable parameters. The predicted $T$ dependence for $T\leq T_{\rm N}$ in Fig.~\ref{fig:Chi}(c) is in reasonable agreement with the measured $\chi(T)$ data (open red circles) plotted in the same figure.  The experimental $\chi(T\to0)$ value in in Fig.~\ref{fig:Chi}(c) is slightly lower than the MFT prediction, which may arise from a slight deviation from a random distribution of the orientation of the grains in the polycrystalline sample.

\begin{figure}
	\includegraphics[width=3.3in]{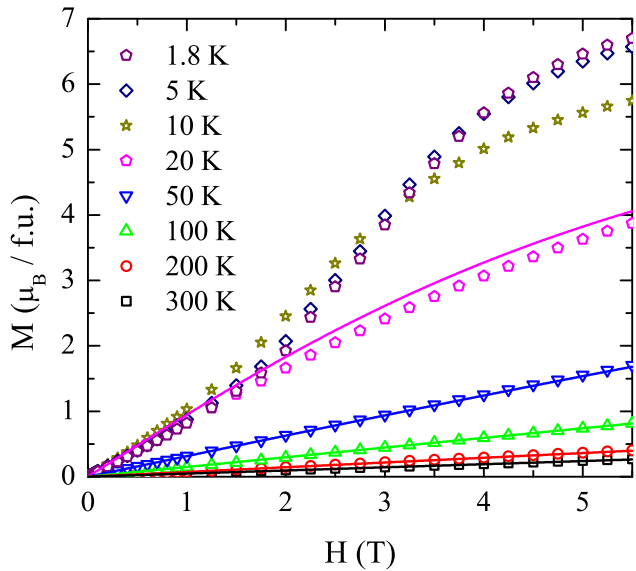}
	\caption{(Color online) Magnetization $M$ versus applied field $H$ isotherms measured at the indicated temperatures for EuNiGe$_3$. Solid curves of corresponding color are the theoretical predictions by MFT in Eq.~(\ref{Eq:MzVsH}) for the data in the paramagnetic state ($T > T_{\rm N} = 13.6$~K). In the ordinate label, f.u.\ means formula unit.  The deviation of the fit from the data for $T=20$~K is due to the presence of short-range AF order at that $T$, and the S-shaped behavior for $T<T_{\rm N}$ is due to occurrence of spin-flop transitions with a distribution of spin flop fields due to the polycrystalline nature of the sample.}
	\label{fig:MH}
\end{figure}

$M$ versus $H$ isotherms were measured for EuNiGe$_3$ in the $H$ range 0--5.5~T at various temperatures as shown in Fig.~\ref{fig:MH}. The MFT prediction per spin in the paramagnetic state at $T>T_{\rm N}$ is\cite{Johnston2011}
\be
\bar{\mu}_z = B_S \left[\frac{3f \bar{\mu}_z }{(S+1)t}+\frac{h}{t}\right],
\label{Eq:MzVsH}
\ee
where $\mu_z$ is the magnetization induced in the direction of the applied field by the applied field, $\bar{\mu}_z\equiv \mu_z/\,\mu_{\rm sat}$ and the reduced applied magnetic field $h$ is defined as $h\equiv g\mu_{\rm B}H/(k_{\rm B}T_{\rm N})$. All of the parameters $g,\ S,\ f$ and~$T_{\rm N}$ in Eq.~(\ref{Eq:MzVsH}) were already determined above.  The $M(H)$ isotherms for $T>T_{\rm N}$ calculated from numerical solution of Eq.~(\ref{Eq:MzVsH}), where $M = N\mu_z$ and $N$ is the number of spins, are compared with no adjustable parameters with the corresponding experimental $M(H)$ data in Fig.~\ref{fig:MH}.  A proportional $M(H)$ behavior is predicted and observed for $T\geq50$~K, whereas negative curvature in $M(H)$ is predicted and observed at 20~K\@. However, the calculated curve for $T=20$~K is slightly above the observed data due to dynamical short-range AF ordering in the sample on approaching $T_{\rm N} = 13.6$~K from above, which suppresses the magnetization.  Such dynamical short-range ordering effects above $T_{\rm N}$ are not taken into account in MFT (see also the next section).  At temperatures below $T_{\rm N}$, the maximum observed magnetization of 6.70~$\mu_{\rm B}$/Eu at $T=1.8$~K and $H=5.5$~T is approaching the saturation moment $\mu_{\rm sat}= 7~\mu_{\rm B}$/f.u.\ expected for Eu$^{+2}$ with $S=7/2$ and $g=2$.

The $M(H)$ isotherms below $T_{\rm N}$ at $T=1.8$, 5 and 10~K show an S-shaped dependence on $H$. Qualitatively, this can be explained by a series of field-induced first-order spin-flop transitions where the ordered moments flop to a perpendicular orientation with respect to the applied field.  In order for a spin flop transition to occur in a collinear AF, some type of  magnetocrystalline anisotropy must be present that aligns the moments along the easy axis at zero field.  The first-order spin flop transition does not occur at a single field as observed in a single crystal with the field along the easy axis because of the random orientations of the crystallites in the polycrystalline sample.  In a polycrystalline sample, one expects the spin flop field in a grain with its easy axis at an angle $\theta$ to the field to obey $H_{\rm flop}(\theta) = H_{\rm flop}(\theta=0)/\cos\theta$.  Thus the spin flop field increases with increasing $\theta$.  When the easy axis is perpendicular to the applied field ($\theta=90^\circ$), a spin flop transition is not possible because the orientation of the ordered moments is already perpendicular (in $H=0$) to the field direction.  From Fig.~\ref{fig:MH}, we infer that $H_{\rm flop}(\theta=0,\, T\to0) \sim 1.5$~T\@.  A calculation within MFT of the powder-averaged $M(H)$ for a polycrystalline sample, incorporating both the anisotropy field and the $\theta$-dependent distribution of spin flop transition fields, is beyond the scope of the present work.

\subsection{\label{HC} Heat capacity measurements}

\begin{figure}
	\includegraphics[width=3.3in]{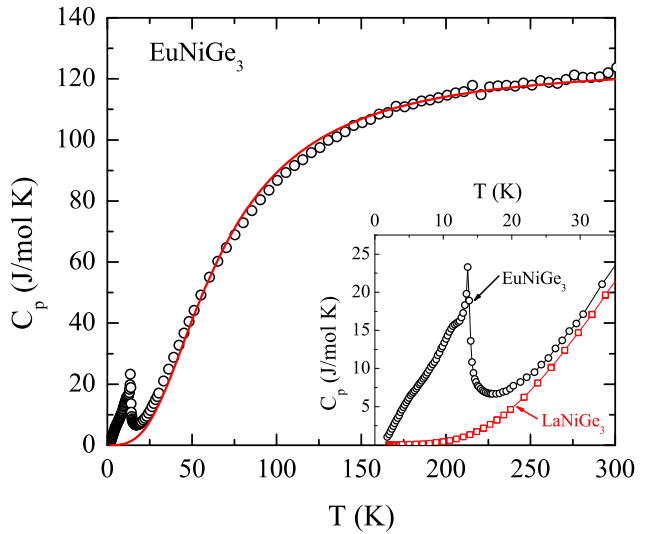}
	\caption{(Color online) Heat capacity $C_{\rm p}$ versus temperature $T$ for EuNiGe$_3$. The red curve is a fit by a Pad\'e approximant of the Debye function in Eq.~(\ref{eq:Debye}) to the data at $T> 25$~K\@. An extrapolation of the fit to $T=0$ is also shown.  Inset: Expanded plot of $C_{\rm p}(T)$ for EuNiGe$_3$ at low~$T$ (black circles), together with background $C_{\rm p}(T)$ data for LaNiGe$_3$ (red squares)\cite{Anand2008} with a renormalized $T$ scale (see text).}
	\label{fig:Cp}
\end{figure}

The $C_{\rm p}$ of EuNiGe$_3$ was measured at $H=0$ in the $T$ range 1.8--300~K and the data are plotted in Fig.~\ref{fig:Cp}. A sharp $\lambda$-shaped peak is observed at $T_{\rm N}=13.6$~K as shown in more detail in the inset of Fig.~\ref{fig:Cp}, confirming that the AF transition observed in the magnetization measurements is a bulk magnetic phase transition. The $C_{\rm p}(300$~K) = 122~J/(mol\,K) is  approaching the classical Dulong-Petit high-$T$ limit $C_{\rm V}=3nR=124.7$~J/(mol~K) for the heat capacity of acoustic lattice vibrations at constant volume, where $n=5$ is the number of atoms per formula unit and $R$ is the molar gas constant.

The Debye model describes the heat capacity versus $T$ due to such lattice vibrations by\cite{Kittel2005}
\be
C_{\rm V}=9 R \left(\frac{T}{\Theta_{\rm D}}\right)^3 \int_0^{\Theta_{\rm D}/T} \frac{x^4 e^x}{(e^x-1)^2}\, dx,
\label{eq:Debye}
\ee 
where $\Theta_{\rm D}$ is the Debye temperature determined from heat capacity measurements. In addition, for a metal one can add a linear $\gamma T$ term to Eq.~(\ref{eq:Debye}) to account for the electronic specific heat contribution, where $\gamma$ is the Sommerfeld electronic specific heat coefficient, and for a magnetic material one can add the magnetic contribution $C_{\rm mag}(T)$.

An accurate analytic Pad\'e approximant of the Debye function that we recently formulated to simplify fitting of experimental $C_{\rm p}(T)$ data by the Debye theory\cite{Goetsch2012} was used in place of Eq.~(\ref{eq:Debye}) to fit our data.  We fitted our $C_{\rm p}(T)$ data from $T=25$ to~300~K because below $\approx 25$~K the magnetic heat capacity contribution $C_{\rm mag}$ becomes significant (see below).  Also, because of the presence of $C_{\rm mag}$, $\gamma$ could not be accurately determined from the $C_{\rm p}(T\to0)$ data. When allowed to vary, it refined to the value 2(2)~mJ/mol\,K$^2$. Therefore, $\gamma$ was fixed at zero for the final fit. The only adjustable parameter in the final fit was $\Theta_{\rm D}$, which was found to be $\Theta_{\rm D}=268(2)$~K\@.  As seen in Fig.~\ref{fig:Cp}, a reasonably good fit of the data by the Debye model is obtained over the entire temperature range above 25~K\@. 

\begin{figure}
	\includegraphics[width=2.6in]{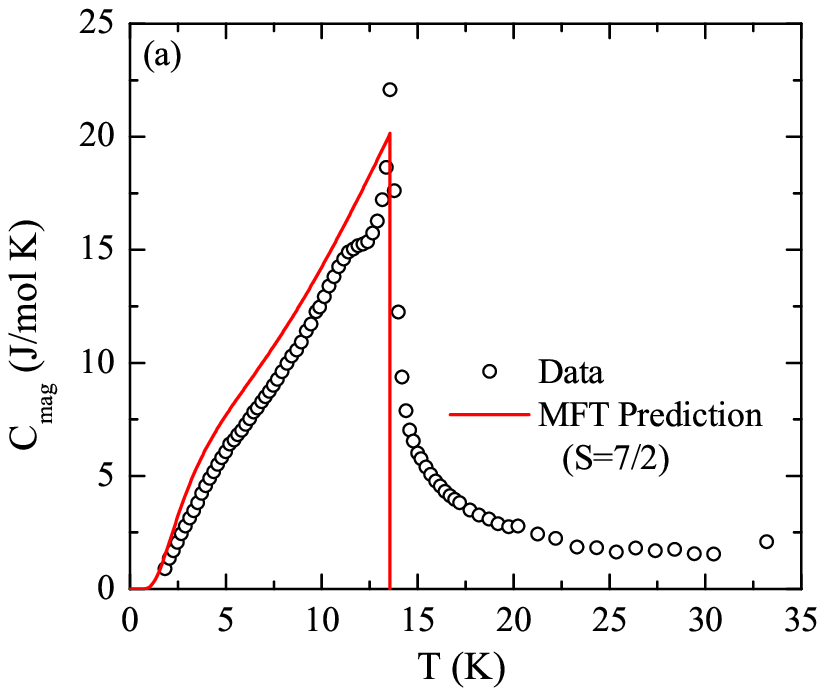}
	\includegraphics[width=2.6in]{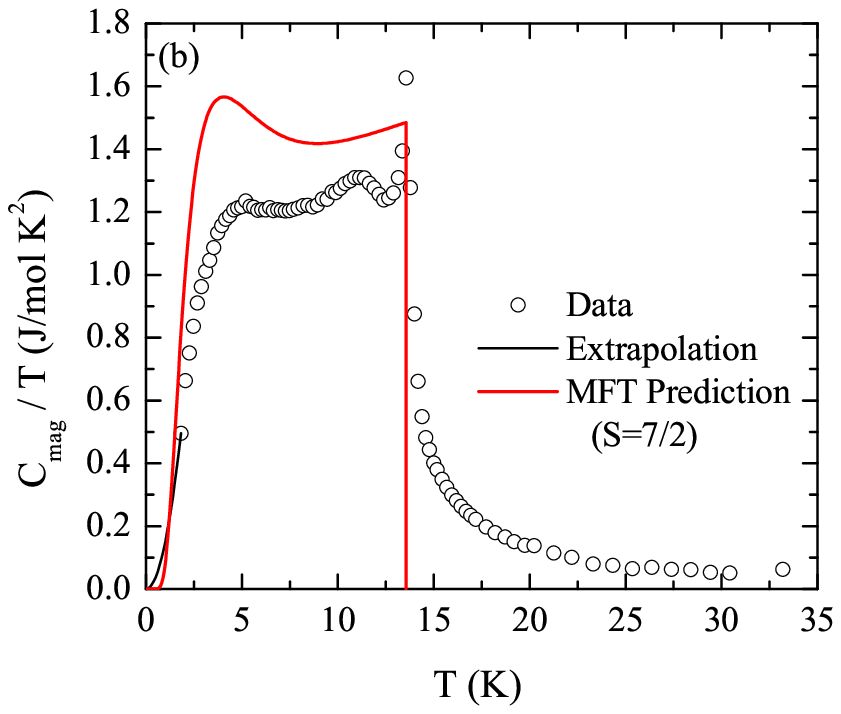}
	\includegraphics[width=2.6in]{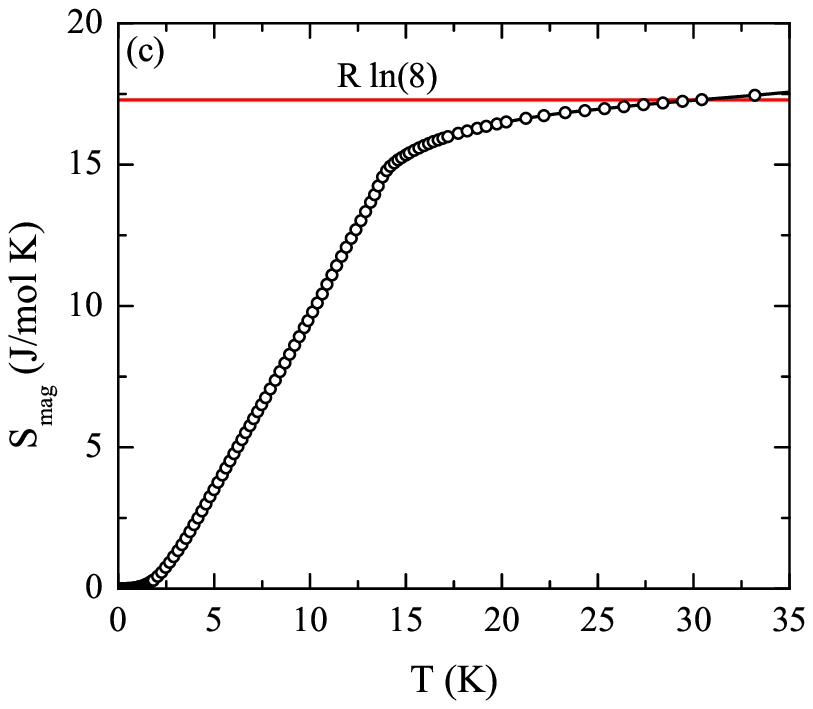}
	\caption{(Color online) (a) Magnetic contribution $C_{\rm mag}$ to the heat capacity of EuNiGe$_3$ versus temperature $T$\@. (b) $C_{\rm mag}/T$ versus $T$. The black curve at low $T$ in (b) is an extrapolation of the data from $T=1.8$~K to $T=0$.  The red curves in (a) and (b) are the predictions of MFT in Eq.~(\ref{eq:Cmag}) for spin $S=7/2$. (c) Magnetic contribution $S_{\rm mag}$ to the entropy versus $T$\@. The horizontal red line is the expected high-$T$ limit $S_{\rm mag} = R\ln(8)$ = 17.29~J/mol\,K for $S=7/2$.}
	\label{fig:Cmag}
\end{figure}

In order to isolate $C_{\rm mag}(T)$, $C_{\rm p}(T)$ data for the isostructural nonmagnetic reference compound LaNiGe$_3$ from Ref.~\onlinecite{Anand2008} was used as a heat capacity background, after correcting for the formula weight (FW) difference by multiplying the temperature scale for $C_{\rm p}(T)$ of LaNiGe$_3$ by ${\rm \sqrt{FW_{LaNiGe_3}/FW_{EuNiGe_3}}}$.  From the inset of Fig.~\ref{fig:Cp}, the renormalized $C_{\rm p}(T)$ for LaNiGe$_3$ is seen to be a reasonable estimate of the background heat capacity of EuNiGe$_3$. The $C_{\rm mag}(T)$ obtained by subtracting the renormalized $C_{\rm p}(T)$ of LaNiGe$_3$ from the $C_{\rm p}(T)$ data for EuNiGe$_3$ is plotted in Fig.~\ref{fig:Cmag}(a) and $C_{\rm mag}(T)/T$ is plotted in Fig.~\ref{fig:Cmag}(b).

MFT predicts that $C_{\rm mag}$ of a Heisenberg spin system containing identical crystallographically equivalent spins, per mole of spins, is\cite{Johnston2011}
\be
\frac{C_{\rm mag}}{R}=-\frac{3S}{S+1} \bar{\mu}_0(t) \frac{d \bar{\mu}_0(t)}{dt}.
\label{eq:Cmag}
\ee
There are no adjustable parameters in this prediction if $S$ and $T_{\rm N}$ are known, as they are here.  Comparisons of the predicted $C_{\rm mag}(T)$ and $C_{\rm mag}(T)/T$ from Eq.~(\ref{eq:Cmag}) with our experimental data are shown in Figs.~\ref{fig:Cmag}(a) and \ref{fig:Cmag}(b), respectively. The hump in the experimental $C_{\rm mag}(T)$ at $T\sim T_{\rm N}/3$, which is much more pronounced in the plot of $C_{\rm mag}(T)/T$, is reproduced by the MFT calculation.  This hump in $C_{\rm mag}(T)$ increases in magnitude as $S$ increases, is particularly noticable for $S=7/2$, and is not visible for, {\it e.g.}, $S=1/2$.\cite{Johnston2011} The hump arises in MFT from the combined effects of the $T$ dependence of the energy splitting of the Zeeman levels of the spin due to the $T$ dependence of the ordered moment and resulting $T$ dependence of the exchange field seen by each moment, together with the $T$ dependence of the Boltzmann populations of those levels.  The hump in $C_{\rm mag}(T)$ observed at $T\sim T_{\rm N}/3$ for magnetically ordered compounds containing Eu$^{+2}$ or Gd$^{+3}$ with $S=7/2$ is sometimes misinterpreted as arising from either an electronic magnetic Schottky anomaly that is combined with a $T^3$ magnon contribution to reproduce the observed $C_{\rm mag}(T)$, as evidence for some type of magnetic phase transition, or as a giant nuclear Schottky anomaly induced by the ordered moments.

The magnetic contribution $S_{\rm mag}(T)$ to the entropy was calculated from the $C_{\rm mag}(T)$ derived from our experiments according to
\be
S_{\rm mag}(T)=\int_0^T \frac{C_{\rm mag}(T)}{T}\, dT.
\ee
Because $C_{\rm mag}(T)$ data were not obtained below 1.8~K, the $C_{\rm mag}/T$ data were extrapolated from $T=1.8$~K to $T=0$ using, for simplicity, the $T^2$ dependence predicted by spin wave theory for a three-dimensional AF in the absence of an anisotropy gap.  The calculated entropy between $T=0$ and 1.8~K on the basis of this extrapolation is 0.30~J/mol K\@.  This is an upper limit since the presence of an anisotropy gap would instead give an exponential decrease in $C_{\rm mag}(T)$ below 1.8~K\@.  As seen in Fig.~\ref{fig:Cmag}(c), the molar $S_{\rm mag}$ saturates to the value $S_{\rm mag}(T\to\infty)\approx R\ln (8)$ expected from quantum statistics according to $S_{\rm mag}(T\to\infty) = R \ln (2S+1)$, where $S=7/2$ for Eu$^{+2}$. Even though the MFT prediction has significant deviations from the measured $C_{\rm mag}/T$ data at $T \leq T_{\rm N}$ in Fig.~\ref{fig:Cmag}(b), the area (magnetic entropy) between the calculated and observed data that is missing below $T_{\rm N}$ is recovered above $T_{\rm N}$. The finite $C_{\rm mag}(T)$ for $T>T_{\rm N}$ arises from dynamical spin correlations at $T>T_{\rm N}$, as often observed, that are not taken into account by MFT\@.  As a result, $S_{\rm mag}(T_{\rm N}=13.6~{\rm K})=14.3$~J/mol~K is 83\% of the asymptotic high-$T$ limit, with the remainder recovered by $\approx 30~{\rm K} = 2.2\,T_{\rm N}$. 

\section{\label{Conclusion} Summary and Conclusions}

\begin{table*}
 \caption{\label{table:properties_comparison} Summary of the physical properties of EuNiGe$_3$. The properties of EuNi$_2$Ge$_2$ are also presented for comparison. Listed are: the tetragonal lattice parameters $a$ and $c$ at room temperature, the N\'eel temperature $T_{\rm N}$, Weiss temperature $\theta_{\rm p}$, Curie constant $C$, effective magnetic moment $\mu_{\rm eff} = \sqrt{8C}$ of the Eu, and Debye temperature determined from heat capacity ($\Theta_{\rm D}$) and resistivity ($\Theta_{\rm R}$) measurements.}
 \begin{ruledtabular}		
		\begin{tabular}{l c c c c c c c c c}
		Compound & $a$  & $c$  & $T_{\rm N}$ & $\theta_{\rm p}$ & $C$  & $\mu_{\rm eff}$  & $\Theta_{\rm D}$ & $\Theta_{\rm R}$ & Ref.	\\ 
& (\AA) & (\AA) & (K) & (K) & (cm$^3$\,K/mol) & ($\mu_{\rm B}$) & (K) & (K)\\\hline
		EuNiGe$_3$ 		& 4.3366(1)	& 9.8802(2) & 13.6 	& 5.3(3)		&  7.66(2)	& 7.83(2)	& 268(2)	& 265(2)	& This work	\\
		EuNi$_2$Ge$_2$ 	&  				&  			& 30.8 	& $-$9.7 	& 			& 7.69 		&  			&			& \onlinecite{Budko1999} 	\\
		EuNi$_2$Ge$_2$	& 4.144(3) 		& 10.15(1) 	& 30 	& $-$8		&  7.7 		&			&			&			& \onlinecite{Felner1978}\\
		\end{tabular}
 \end{ruledtabular}
\end{table*}

A nearly single-phase polycrystalline sample of EuNiGe$_3$ was synthesized and its physical properties were investigated. Rietveld refinements of the powder XRD data confirmed that this compound crystallizes in the body-centered-tetragonal BaNiSn$_3$-type structure with space group $I4mm$ as previously reported.  The $\rho$, $\chi$ and $C_{\rm p}$ measurements consistently reveal an AF ordering transition at $T_{\rm N}=13.6$~K\@. A summary of some of the results from these measurements is given in Table~\ref{table:properties_comparison}.

The $\rho(T)$ measurements of EuNiGe$_3$ reveal metallic behavior. The large ${\rm RRR = 54.8}$ indicates the high quality of the sample. The data for $T>T_{\rm N}$ are well-described by the Bloch-Gr\"uneisen theory for the $T$-dependent resistivity arising from electron-phonon scattering. A fit to the data for $T_{\rm N}< T < 350$~K by the theory yielded a Debye temperature $\Theta_{\rm R}=265(2)$~K\@. The $\rho$ decreases rapidly on cooling below $T_{\rm N}$ due to loss of spin-disorder scattering.  Since the compound is metallic, the magnetic coupling between the Eu spins likely arises mainly from the indirect RKKY interaction mediated by the conduction electrons.

The $M(H,T)$ and $\chi(T)$ measurements of EuNiGe$_3$ showed the presence of long-range AF order in this system at $T_{\rm N} \approx 14$~K\@. A fit of $\chi^{-1}(T)$ by the Curie-Weiss law at $T \geq 50$~K revealed a Curie constant consistent with the presence of Eu$^{+2}$ ions with $S=7/2$ and $g=2$, and a positive Weiss temperature $\theta_{\rm p} =5.3(3)$~K, indicating that ferromagnetic interactions are dominant despite the occurrence of long-range AF ordering. There was no evidence from our measurements that the Ni atoms are magnetic.  A low-field $\chi(T)$ measurement at $T\leq T_{\rm N}$ was compared with our prediction\cite{Johnston2012} from MFT for the polycrystalline average of the anisotropic $\chi(T)$ of a collinear antiferromagnet below its N\'eel temperature and good agreement was found. Carrying out such a fit for a polycrystalline AF was one of the goals of this work as discussed in the introduction.   A field-induced spin flop transition was inferred from the S-shaped $M(H)$ curves at $T <T_{\rm N}$, with a low-$T$  onset field of $H \sim 1.5$~T that was spread out to higher fields due to the polycrystalline nature of the sample.

The presence of dominant FM interactions in EuNiGe$_3$ that orders antiferromagnetically led us to propose that these interactions are between Eu spins within the $ab$-plane, with subdominant AF interactions between spins in adjacent planes along the $c$-axis.  From these interactions, we propose that the collinear AF structure is A-type, in which the Eu spins within an $ab$-plane in Fig.~\ref{fig:structure}(a) are aligned ferromagnetically with respect to each other and the spins in adjacent layers along the $c$-axis are aligned antiferromagnetically to each other.  The most likely ordered moment axis is either the $c$-axis or an axis in the $ab$-plane.  If it is the $c$-axis, magnetocrystalline anisotropy effects would presumably not cause a distortion of the crystal structure on cooling below $T_{\rm N}$.  However, if the ordered moments are in the $ab$-plane, the A-type collinear ordering breaks the fourfold rotational symmetry about the $c$ axis of the tetragonal room-temperature crystal structure, and an orthorhombic crystal distortion may be expected to occur on cooling below $T_{\rm N}$. 

Our $C_{\rm p}(T)$ data were fitted by the Debye model from $T=25$~K to 300~K, yielding a Debye temperature $\Theta_{\rm D}=268(2)$~K\@. This value is the same within the error bars as the Debye temperature $\Theta_{\rm R}=265(2)$~K determined from the $\rho(T)$ measurements, a rare occurrence.\cite{Goetsch2012}  The data exhibited a sharp $\lambda$-shaped peak at $T_{\rm N}$, which allowed the precise value of the N\'eel temperature to be determined to be $T_{\rm N} = 13.6$~K\@.  The magnetic heat capacity contribution $C_{\rm mag}(T)$ and the magnetic entropy $S_{\rm mag}(T)$ were extracted and analyzed by MFT\@.  The high-$T$ limiting value of the entropy $S_{\rm mag}\approx R\ln(8) = R\ln(2S+1)$ is consistent with our $\chi(T>T_{\rm N})$ data that indicated $S=7/2$.  Significant short-range AF correlations occur above $T_{\rm N}$, with about 83\% of the maximum magnetic entropy present at $T_{\rm N}$ and the remaining 17\% recovered by $\approx 30\ {\rm K} = 2.2\, T_{\rm N}$\@. 

As shown in Fig.~\ref{fig:structure}, EuNiGe$_3$ crystallizes in the BaNiSn$_3$-type structure which is similar to the ThCr$_2$Si$_2$-type structure. In fact, the compound EuNi$_2$Ge$_2$ crystallizes in the latter structure and its properties have been measured.\cite{Budko1999,Felner1978}  The Eu sublattices in the two compounds are identical and the Ni atoms in both compounds are believed to be nonmagnetic.  Therefore, we compare some properties of these two compounds in Table~\ref{table:properties_comparison}.  Both compounds contain Eu$^{+2}$ ions with $S=7/2$ and $g=2$ and order antiferromagnetically with EuNiGe$_3$ having the lower $T_{\rm N}$. A significant difference between these compounds is the positive $\theta_{\rm p}$ in EuNiGe$_3$, indicating dominant FM interactions as discussed above, and a negative one in EuNi$_2$Ge$_2$ indicating dominant AF interactions. This difference indicates that the magnetic interactions between the Eu spins are quite different in the two compounds and therefore suggests that the resultant AF structures may also be different.  As discussed above, we propose that EuNiGe$_3$ has an A-type AF structure.  The anisotropic $\chi(T)$ measurements on EuNi$_2$Ge$_2$ single crystals\cite{Budko1999} suggest that the ordered moments lie in the $ab$-plane with a collinear AF structure and multiple AF domains. Alternatively, a comparison of our recent predictions\cite{Johnston2012} of the anisotropic $\chi(T)$ of planar noncollinear AFs with the $\chi(T)$ data\cite{Budko1999} for EuNi$_2$Ge$_2$ suggests that this compound may have a planar noncollinear AF structure with the ordered moments aligned within the $ab$-plane.  It would be useful and interesting to determine the AF structures of both EuNiGe$_3$ and EuNi$_2$Ge$_2$ by magnetic neutron or x-ray scattering measurements and to correlate the results with the respective $\chi(T)$ data for these compounds.

\acknowledgments

This research was supported by the U.S. Department of Energy, Office of Basic Energy Sciences, Division of Materials Sciences and Engineering. ÊAmes Laboratory is operated for the U.S. Department of Energy by Iowa State University under Contract No.~DE-AC02-07CH11358.

\end{document}